\documentclass[%
 aip,amsmath,amssymb,floatfix,
 reprint,
 jcp
]{revtex4-2}

\usepackage{graphicx}% Include figure files
\usepackage{dcolumn}% Align table columns on decimal point
\usepackage{bm}
\usepackage{comment}
\usepackage{hyperref}% add hypertext capabilities
\usepackage[T1]{fontenc}
\usepackage{color}
\usepackage{xkcdcolors}
\usepackage{soul}

\begin{document}

\title{What is the best simulation approach for measuring local density fluctuations near solvo/hydrophobes?}

\author{Nigel B. Wilding}
\author{Robert Evans}
\author{Francesco Turci}
\affiliation{ H.H.~Wills Physics Laboratory, University of Bristol, Royal Fort,  Bristol BS8 1TL, U.K.}

\date{\today}

\begin{abstract}

Measurements of local density fluctuations are crucial to characterizing the interfacial properties of equilibrium fluids. A specific case that has been well-explored involves the heightened compressibility of water near hydrophobic entities. Commonly, a spatial profile of local fluctuation strength is constructed from measurements of the mean and variance of solvent particle number fluctuations in a set of contiguous sub-volumes of the system adjacent to the solvo/hydrophobe. An alternative measure proposed by Evans and Stewart (J. Phys.: Condens. Matter {\bf 27} 194111 (2015)) defines a local compressibility profile in terms of the chemical potential derivative of the spatial number density profile. Using Grand Canonical Monte Carlo simulation, we compare and contrast the efficacy of these two approaches for a Lennard-Jones solvent at spherical and planar solvophobic interfaces, and SPC/E water at a hydrophobic spherical solute. Our principal findings are that: (i) the local compressibility profile  $\chi({\bf r})$ of Evans and Stewart is considerably more sensitive to variations in the strength of local density fluctuations than the spatial fluctuation profile ${\cal F}({\bf r})$ and can resolve much more detailed structure; (ii) while the local compressibility profile is essentially independent of the choice of spatial discretization used to construct the profile, the spatial fluctuation profile exhibits strong systematic dependence on the size of the subvolumes on which the profile is defined. We clarify the origin and nature of this finite-size effect. 

\end{abstract}

\maketitle

\section{Introduction}
\label{sec:intro}

Fluids adsorbed at substrates or extended solutes typically display distinctive local density profiles together with enhanced density fluctuations. Perhaps the best-known examples arise in the context of the continuous wetting and drying surface phase transitions that occur for a planar substrate of macroscopic interfacial area. A continuous wetting transition corresponds to adsorption from the bulk vapor whereby a film of liquid intrudes between the vapor and a (sufficiently attractive) substrate whose thickness -- and hence the Gibbs excess adsorption --  diverges continuously as a thermodynamic control parameter is varied. The approach to the transition is accompanied by growing density-density correlations parallel to the substrate, arising from capillary wave-like fluctuations, whose correlation length $\xi_\parallel$ diverges, see e.g. the reviews by Dietrich\cite{Dietrich1988} and Bonn et.\;al\cite{Bonn:2009if}. Drying is the analogue of wetting, but now the bulk is liquid, and a film of vapor intrudes between it and the (repulsive or very weakly attractive) substrate. Recent theoretical and numerical studies by Evans {\em et. al.}\cite{EvansStewartWilding2017,EvansStewartWilding2019} have shown that for realistic choices of fluid-fluid and substrate-fluid potentials a continuous drying transition is likely to occur at bulk coexistence by decreasing the strength of the attractive interaction between the substrate and the fluid. At a continuous drying transition, the thickness of the film of vapor and the negative of the Gibbs excess adsorption diverge, accompanied by growing density-density correlations characterized by a diverging correlation length $\xi_\parallel$. Other situations where a diverging parallel correlation length occurs are: i) near a pre-wetting critical point, ii) a fluid confined between two identical parallel planar substrates on approaching a capillary critical point \cite{EvansParry1990} and ii) a fluid in an asymmetric planar slit where one substrate prefers to wet and the other prefers to dry, e.g. Stewart and Evans\cite{StewartEvans2012} and references therein. For all these cases there is ample understanding of the nature of the surface criticality to enable proper analysis of simulation results. 

The situation is different for adsorption at solutes of finite extent; the optimal way of measuring solvent density fluctuations is not immediately obvious. A suitable measure should treat the full range of solute size and shape; solutes might range from simple near spherical molecules to much larger objects such as colloidal particles and include more complex entities such as proteins or biological molecules. Often, the solvent considered is liquid water, and then one is concerned with hydrophobic solutes. Various attempts have been made to connect enhanced density fluctuations in water with the degree of hydrophobicity and also with drying phenomena; the recent review of Rego and Patel\cite{rego2022} provides many references.  Of course, one cannot define a parallel correlation length for density-density correlations near a general solute. Rather, one requires a robust measure that describes how the strength of density fluctuations varies with the distance from the solute,  the size of the solute, and the proximity of the thermodynamic state point of the solvent to bulk coexistence.

Experience with planar substrates indicates that there is a natural measure. This is the local compressibility $\chi({\bf r})$ defined as the derivative of the equilibrium density profile $\rho({\bf r})$ with respect to (w.r.t) the chemical potential $\mu$, ie. $\chi({\bf r})=\partial\rho({\bf r})/\partial \mu|_T$,  which in bulk, where the density is constant, is proportional to the usual thermodynamic quantity, the isothermal compressibility. In planar substrate geometry, the maximum of $\chi(z)$ diverges (essentially) in the same way as $\xi_\parallel^2$ on approaching a continuous surface transition. Employing classical Density Functional Theory (DFT) calculations for Lennard-Jones (LJ) fluids, Evans and Stewart\cite{EvansStewart2015} demonstrated that $\chi(z)$ provides a valuable measure for characterizing hydrophobicity, or more generally solvophobicity, at planar substrates. Specifically, they showed that for substrates where Young’s contact angle $\theta$ is very large, but $\theta< \pi$, the maximum in $\chi(z)$ is at least an order of magnitude larger than the bulk value $\chi_b$ and occurs at distances $z$ within one or two atomic diameters of the substrate. They argued that $\chi(z)$ is a much sharper indicator of the degree of solvophobicity of a substrate, as measured by the contact angle, than is the extent of density depletion, as extracted from the density profile $\rho(z)$. Subsequent papers reinforced these ideas using Grand Canonical Monte Carlo (GCMC) simulations and DFT for LJ fluids\cite{EvansStewartWilding2016,EvansStewartWilding2017}  and simulations of SPC/E water\cite{EvansWilding2015}. Later studies by Eckert et.al.\cite{EckertSchmidt2020,Eckert_2023} introduced a (closely related) local thermal susceptibility $\chi_T(z)=\partial\rho( z)/\partial T|_\mu$ that is the temperature derivative of the density profile. Coe {\em et. al.}\cite{CoeEvansWildingPRE2022} showed that on approaching critical drying, the singular behavior of $\chi(z)$ drives identical singular behavior in $\chi_T(z)$. 

Note that in an early paper \cite{StewartEvans2012}, and references therein,  $\chi(z)$ was termed the \textit{local susceptibility} since it is the direct analogue for continuum fluids of the layer magnetic susceptibility $\chi_n$ in an Ising lattice subject to a surface magnetic field. Specifically, $\chi_n=\partial m_n/\partial h$ is the derivative of the (average) magnetization $m_n$, in the $n$-th layer away from the surface, w.r.t.~the external magnetic field $h$; recall that $h$ is equivalent to the chemical potential $\mu$. The layer susceptibility provides a powerful measure of the strength of local magnetization fluctuations in layer $n$ and measurements of $\chi_n$, using Monte Carlo simulations, have played a key role in elucidating the fundamental physics of wetting transitions and phase transitions arising from confinement\cite{Binder1995,Binder2003}. 

Arguments that $\chi(r)$ provides an effective measure of the degree of solvo/hydrophobicity at finite spherical solutes were presented in two recent papers\cite{CoeEvansWildingPRL2022,CoeEvansWildingJCP2023} which draw upon ideas from earlier studies of critical drying at (very large) spherical particles\cite{StewartEvans2005}. Similarly to planar systems, a pronounced peak develops in $\chi(r)$ close to the solute whose height increases with the proximity to bulk coexistence and now also with the radius of the solute; the latter acts as a further variable in a comprehensive scaling analysis of solvent thermodynamics and density fluctuations at a solvophobe. Importantly, these studies, together with those for the planar cases, emphasize the lack of water-specific mechanisms for the behavior near an extended hydrophobe: the physics of the density fluctuations around a generic solvophobe should be the same.

Of course, there are other measures of density fluctuations near hydrophobes and Evans and Stewart\cite{EvansStewart2015} provide a summary. For models of water, or any fluid,  confined in a slit pore, the mean square fluctuation (variance) of the total particle number provides a valuable measure of the overall compressibility of the system and is often investigated in GCE simulation, see the commentary by Bratko\cite{Bratko:2010} and references therein. Here we focus on\textit{ local} measures.  Acharya {\em et.\:al.} \cite{AcharyaGarde2010} (see also Sarupria and Garde\cite{SarupriaGarde2009} and the review by  Jamadagni {\em et.\:al.} \cite{Jamadagni2011}) attempted a definition of a local compressibility that involves a derivative of the density profile $\rho(z)$ w.r.t.~pressure. However, it is not clear precisely what the pressure is and how the derivative is performed. 

In their paper, Acharya {\em et. al.} \cite{AcharyaGarde2010} also describe another quantity, denoted as $\chi_{fl}(z)$, which measures the {\em variance} of particle number fluctuations in a slab of certain thickness $\Delta z$ located at a distance $z$ from a substrate. This quantity provides a simple means of defining a spatial fluctuation profile and is essentially the quantity ${\cal F}(z)$ that we shall define in Sec.~\ref{sec:localfluctprof}.  Using molecular dynamics simulations, Willard and Chandler\cite{WillardChandler2014} measured ${\cal F}(z)$ for SPC/E water near a planar Lennard-Jones $12$-$6$ wall and found that this quantity increased, for $z$ close to the wall, with decreasing wall-fluid attraction, i.e. with increasing contact angle. They argued that ${\cal F}(z)$ provides a quantitative measure of the degree of hydrophobicity. Evans and Stewart\cite{EvansStewart2015} noted that their DFT results for $\chi(z)$ in the LJ fluid exhibited similar trends to those displayed in the SPC/E results for ${\cal F}(z)$, but could not perform direct comparisons. 

In the present work, we perform a systematic comparison of the utility of the spatial fluctuation profile ${\cal F}({\bf r})$ (i.e. the generalization to arbitrary geometry of ${\cal F}(z)$), and the local compressibility profile $\chi({\bf r})$, for probing local density fluctuations. Our paper is arranged as follows. Sec.~\ref{sec:background} describes some pertinent background to fluctuations in particle number. We consider the definitions of  ${\cal F}({\bf r})$ and $\chi({\bf r})$, emphasizing the distinction between these; $\chi({\bf r})$ is proportional to the correlator (covariance) of the local particle number $N({\bf r})$ and the total particle number $N$ whereas ${\cal F}({\bf r})$  measures the correlator of $N({\bf r})$  with itself.  In Sec.\ref{sec:MCresults} we present GCMC results for three different physical systems: i) a LJ liquid confined between two planar solvophobic walls, ii) a LJ liquid at a spherical solvophobe, and iii) SPC/E water at a spherical hydrophobe. In each case, we compare the relative merits of the two profile measures in quantifying the strength and range of density fluctuations. We conclude in Sec.\ref{sec:discussion} with a summary and discussion.

\section{Background and methods}
\label{sec:background}

\subsection{Spatial fluctuation profile}
\label{sec:localfluctprof}

We start by considering the isothermal compressibility of a bulk (uniform) fluid of volume $V=L^d$ (with $d$ the dimensionality of the system) defined by $\kappa_T=-\frac{1}{V}\left(\frac{\partial V}{\partial p}\right)_T$, with $p$ the system pressure. Within the grand canonical (constant $\mu VT$) ensemble (GCE), $\kappa_T$ can be related to the configurational average $\langle N\rangle $ and variance $\langle N^2\rangle -\langle N\rangle^2$ of the fluctuating total number of particles $N$ by the well-known expression~\cite{HansenMcDonald}:

\begin{equation}
    \kappa_T = \frac{1}{k_BT\rho_b}\frac{\langle N^2\rangle-\langle N\rangle^2}{\langle N\rangle }  \:,
    \label{eq:kappaT}
\end{equation}
 where $\rho_b\equiv \langle \rho\rangle=\langle N\rangle/V$ is the fixed bulk number density. Here we note that since $\langle N\rangle$ scales linearly with $V$, then for $\kappa_T$ to be intensive requires that $\langle N^2\rangle-\langle N\rangle^2$ scales similarly. This is equivalent to requiring that the variance of the probability density function (pdf) of the number density  $\langle \rho^2\rangle-\langle \rho\rangle^2=k_BT\rho_b^2\kappa_T/V$, which is the standard result \cite{Rovere_1990} for the finite-size scaling of Gaussian density fluctuations that arise as a consequence of the central limit theorem. 

\begin{comment}
Eq.~\ref{eq:kappaT} can be recast in terms of the fluctuations in the number density fluctuations: 
\begin{equation}
  \kappa_T= \frac{1}{k_BT\rho_b}\frac{V(\langle \rho^2\rangle-\langle \rho\rangle^2)}{\langle \rho\rangle} 
\end{equation}
where in the 
\end{comment}
For statistical ensembles in which the total particle number $N$ is fixed, such as the constant-$NVT$ and constant-$NPT$ ensembles, it is clear that  Eq.(~\ref{eq:kappaT}) cannot be applied at the level of the total system. However, an estimate of $\kappa_T$ can potentially be obtained by evaluating Eq.~(\ref{eq:kappaT}) over a domain or `subvolume'\cite{Rovere_1990}. In the limit in which the size of the subvolume $v$ is large on the scale of the particle size, but small compared to the total size of the system so that $v/V\to 0$, fluctuations in the number of subvolume particles $N_v$ will effectively be grand canonical in form. This observation has motivated several authors to attempt to evaluate via constant-$N$ molecular dynamics simulations the spatial dependence of the strength of density fluctuations in inhomogeneous fluids at substrates or solutes by applying Eq.~(\ref{eq:kappaT}) to each of a set of contiguous subvolumes that cover the spatial region of interest\cite{WillardChandler2014,AcharyaGarde2010,SarupriaGarde2009,Turci2023partial}. Below we consider how this strategy is implemented in practice for various solute/solvent geometries of interest. 

As a first example, consider the case of a three-dimensional ($d=3$) slit geometry in which a solvent occupies the space between a pair of planar substrates located at $z=0$ and $z=L$, with the system assumed periodic in the $x$-$y$ plane so that the density profile varies only in the $z$ direction. It is natural to discretize the system in the $z$ direction into identical thin parallel slabs of thickness $\Delta z$, each having equal subvolume $V(z)=L^2\Delta z$, where we have chosen the parallel area=$L^2$. Evaluating the mean and variance of the solvent particle number fluctuations in each slab yields a histogram of the fluctuation profile: 

\begin{equation}
  {\cal F}(z) = \frac{\langle N^2(z)\rangle-\langle N(z)\rangle^2}{\langle N(z)\rangle } \:.
  \label{eq:fluct_planar}
\end{equation}
Here we note in analogy to the discussion for $\kappa_T$ above, that since $\langle N(z)\rangle\sim V(z)$, the numerator of Eq.~\ref{eq:fluct_planar} should similarly scale linearly with $V(z)$ for ${\cal F}(z)$ to provide an intensive measure of the local fluctuations.

Another common scenario considers an extended spherical solute particle fixed at the origin and immersed in a solvent. For this geometry, it is natural to discretize the space surrounding the solute into contiguous spherical shells (concentric with the origin), each of which encompasses the space between some radius $r$ and $r+\Delta r$ and has a subvolume $V(r)=4\pi r^2\Delta r$. Evaluating the mean and variance of the solvent particle number fluctuations within each shell yields a histogram of the radial fluctuation profile:

\begin{equation}
  {\cal F}(r) = \frac{\langle N^2(r)\rangle-\langle N(r)\rangle^2}{\langle N(r)\rangle } \:,
    \label{eq:fluct_sphere}
\end{equation} 
for which similar scaling considerations as for $V(z)$ apply now with regard to $V(r)$.

The approach can readily be extended to the case of an irregular solute, such as a protein molecule, that lacks the symmetry of the above examples. Here, the spatial variation of the fluctuation of the solvent density in $d=3$ can be mapped using a fluctuation profile ${\cal F}({\bf r})$ calculated w.r.t.~a position vector ${\bf r}$ which we assume can range over the discrete lattice vectors of a space-filling structure of eg.~cubic subcells of volume $V({\bf r})=(\Delta l)^3$. Evaluating $\langle N({\bf r})\rangle$ and $\langle N^2({\bf r})\rangle $ for fluctuations in the solvent particle number in each subcell yields a histogram ${\cal F}({\bf r})$:

\begin{equation}
  {\cal F}({\bf r}) = \frac{\langle N^2({\bf r})\rangle-\langle N({\bf r})\rangle^2}{\langle N({\bf r})\rangle } \:.
    \label{eq:fluct_field}
\end{equation} 

Each of the above approaches to defining a discretized fluctuation profile entails a choice for the subvolume size and shape. The shape may be suggested by the geometry of the problem, but the size needs to be sufficiently small to resolve the pertinent features of the local density fluctuations. Furthermore, when operating in a fixed $N$ ensemble, the subvolume should be much smaller than the system size; otherwise, density fluctuations will be suppressed, leading to biased results. However, it transpires that if one chooses a subvolume that is smaller than some multiple of the diameter of the solvent fluid particles, then the results can also be heavily biased.  Aspects of the issues of measuring fluctuations via subvolumes have previously been highlighted by Villamaina and Trizac \cite{Villamaina_2014} and Rom\'{a}n et al \cite{Roman1998} who considered the extent to which Eq.~(\ref{eq:kappaT}) when applied to a square subvolume of a two-dimensional uniform fluid in the constant-$NVT$ ensemble yields an accurate estimate for $\kappa_T$. It was found that it failed to do so for subvolumes less than about $10$ particle diameters in linear size even in the limit when this size was much smaller than that of the system as a whole~\cite{Villamaina_2014}. Related studies of uniform three-dimensional fluids have considered whether one can correct measurements of subvolume compressibility to yield estimates of the bulk compressibility~\cite{SALACUSE2008,Heidari2018,Sevilla2022}. 

The finding that subvolume estimates of $\kappa_T$ can exhibit serious finite-size effects is a consequence of the central limit theorem: when the subvolume size is insufficiently large, the local density fluctuations deviate from the Gaussian form that pertains to very large subvolumes. This is true even for situations that are far removed from a bulk or surface critical point so that the correlation length for density-density fluctuations remains of order the particle size.  The consequence for fluctuation profiles is that the numerator in each of Eqs.~(\ref{eq:fluct_planar}-\ref{eq:fluct_field}) scales nonlinearly with the subvolume, thus engendering a subvolume size dependence of the fluctuation profile. However, to date, the detailed consequences of this subvolume finite-size effect for the sensitivity, resolution, and accuracy of measurements of local density fluctuations in inhomogeneous fluids have not been investigated. We address some of these here.

\subsection{Local compressibility profile}
\label{sec:LCPdef}

The local compressibility profile as introduced by Evans and Stewart~\cite{EvansStewart2015} is defined within the GCE and takes the form

\begin{equation}
   \chi({\bf r})=\left.\frac{\partial \langle \rho({\bf r})\rangle} {\partial \mu}\right|_T \:.
   \label{eq:lc_def}
\end{equation}
Here $\langle\rho({\bf r})\rangle$ is the ensemble average of the instantaneous  density profile $\rho({\bf r})=\sum_{i=1}^N \delta({\bf r}-{\bf r}_i)$, where ${\bf r}$ is the d-dimensional position vector and $\mu$ is the system chemical potential. It is straightforward to show (see appendix) that for a bulk system for which $\chi({\bf r})=\chi_b$ is constant by translational invariance, the local compressibility profile is related to the bulk isothermal compressibility via $\chi_b=\rho_b^2\kappa_T$.

The definition eq.~\ref{eq:lc_def} is formally for a continuous $d$-dimensional density profile.  However, in practice $\chi({\bf r})$ is accumulated as a histogram by discretizing the space of interest into a set of subvolumes of interest. Then ${\bf r}$ belongs to a discrete set of vectors running over the subvolumes, which may, for example, be cubic cells, spherical shells, or planar slabs as outlined in Sec.~\ref{sec:localfluctprof}. In the latter two cases, one obtains one-dimensional profiles $\chi(r)$ and $\chi(z)$ respectively.   

In simulations $\langle \rho({\bf r})\rangle$ is calculated as a configurational average of histograms  $\rho({\bf r})=N({\bf r})/V({\bf r})$ obtained by binning the solvent particle positions into the set of subvolumes\cite{EvansWilding2015,EvansStewartWilding2017} . A convenient approach for obtaining $\chi({\bf r})$ is to explicitly perform the numerical derivative as per Eq.~(\ref{eq:lc_def}), by taking a finite difference. The apparent need to perform two simulations at different $\mu$ to achieve this can be neatly avoided by employing histogram reweighting \cite{FerrenbergSwendsen1988} of a single simulation. In short, a GCE simulation is performed at some chemical potential $\mu$ of interest from which one accumulates an uncorrelated sequence of $M$ simultaneous measurements of $\rho({\bf r})$ and the total number of particles $N$. From this sequence, one first forms an estimate for the ensemble-averaged density profile: 

\begin{equation}
    \langle \rho({\bf r}|\mu)\rangle=\frac{1}{M}\sum_{i=1}^M\rho_i({\bf r}|\mu)\;.
\end{equation}
Then this same sequence of measurements is reweighted w.r.t.~a small notional change $\Delta\mu$ in the chemical potential to yield an estimate for the average profile corresponding to $\mu+\Delta \mu$:

\begin{equation}
    \langle \rho({\bf r}|\mu+\Delta \mu)\rangle=\frac{\sum_{i=1}^M\rho_i({\bf r}|\mu)e^{\beta\Delta\mu N_i}}{\sum_{i=1}^Me^{\beta\Delta\mu N_i}}\:,
\end{equation}
where $N_i$ is the total number of particles for measurement $i$. Since $\Delta\mu$ is notional, it can be chosen to be arbitrarily small (we use $\Delta\mu=10^{-4}$). The local compressibility profile follows simply as
\begin{equation}
    \chi({\bf r}|\mu)= \frac{\langle \rho({\bf r}|\mu+\Delta \mu)\rangle- \langle \rho({\bf r}|\mu)\rangle}{\Delta \mu}\:.
\end{equation}

A straightforward derivation starting from the grand partition function (see the appendix) shows that $\chi({\bf r})$ can also be expressed in terms of a correlator\cite{EvansWilding2015,EvansStewartWilding2017} which takes the form
\begin{eqnarray}
    k_BT\chi({\bf r}) &= &  \langle \rho({\bf r})N\rangle- \langle\rho({\bf r})\rangle\langle N\rangle\\[2mm]    \label{eq:lc_corr}
    &= &  \frac{\left[\langle N({\bf r})  N \rangle-\langle N({\bf r})\rangle \langle N\rangle\right]}{V({\bf r})}\:.
    \label{eq:lc_corr2}
\end{eqnarray}
This expression provides an alternative route to that just described for measuring $\chi({\bf r})$. Furthermore, it serves to expose the similarities and differences between the two approaches for probing local fluctuations represented by the quantities $\chi({\bf r})$ and ${\cal F}({\bf r})$. Comparing Eqs.~(\ref{eq:lc_corr}) and (\ref{eq:fluct_field}) shows that $\chi({\bf r})$ and  ${\cal F}({\bf r})$ are fundamentally distinct in character. Specifically,  $\chi({\bf r})$ correlates the  instantaneous density profile $\rho({\bf r})$, or subvolume particle number $N({\bf r})$, with the instantaneous total particle number $N$, while ${\cal F}({\bf r})$ correlates the subvolume particle number with itself. Thus, the numerical value of the product $\langle N({\bf r})N\rangle$ far exceeds that of $\langle N^2({\bf r})\rangle$. Consequently the central limit theorem (and, by extension, the linear scaling with $V({\bf r})$ at fixed $V$ of the numerator in Eq.~\ref{eq:lc_corr2}) is satisfied to a much greater degree for $\chi({\bf r})$ than is the case for ${\cal F}({\bf r})$, eq.~\ref{eq:fluct_field}. As we shall see, this means that $\chi({\bf r})$ produces the bulk compressibility value even for a minimal choice of the subvolume size on which it is defined, in sharp contrast to ${\cal F}({\bf r})$. 

 We end this Section by noting that the integral of the difference ($\chi({\bf r})-\chi_b$) over the volume available to the fluid measures the surface excess compressibility $\chi_e$, which is the derivative of the Gibbs excess adsorption w.r.t.~chemical potential. $\chi_e$ is proportional to the difference between the variance of the {\em total} number of particles in the inhomogeneous fluid and the corresponding variance in the bulk fluid at the same chemical potential. This quantity provides a powerful measure of the {\em integrated} strength of density fluctuations as demonstrated explicitly for planar systems\cite{EvansStewart2015}.

\section{Monte Carlo simulation results}
\label{sec:MCresults}
 We have investigated the relative merits of the spatial fluctuation profile ${\cal F}$ and the local compressibility $\chi$ via Monte Carlo simulations of three distinct physical setups:

\begin{enumerate}
\item[(i)] A Lennard-Jones solvent confined by a pair of solvophobic planar walls; the profiles $\chi(z)$ and ${\cal F}(z)$ normal to the walls are studied using planar slab subvolumes.

\item[(ii)] A Lennard-Jones solvent in contact with a solvophobic spherical solute;  the radial profiles $\chi(r)$ and ${\cal F}(r)$ are studied using spherical shell subvolumes.

 \item[(iii)] SPC/E water~\cite{Berendsen:1987aa} in contact with a hydrophobic spherical solute; the profiles $\chi({\bf r})$ and ${\cal F}({\bf r})$ are studied using cubic subvolumes.
\end{enumerate}

In each instance, we work within the GCE \cite{FrenkelSmit}, employing the venerable algorithm of Metropolis, Rosenbluth, Rosenbluth, Teller, and Teller\cite{Metropolis1953}, which is celebrated in this special issue. 

\subsection{Fluctuations in a slit with solvophobic planar walls investigated with planar slab subvolumes}
\label{sec:slitcomparison}
 We construct a slit geometry from a cubic simulation box of volume $L^3$ by placing planar walls at $z=0$ and $z=L$, with periodic boundary conditions in the $x$ and $y$ directions. For the solvent, we employ a $12$-$6$ Lennard-Jones (LJ) fluid that is truncated at $r_c=2.5\sigma$ (where $\sigma$ is the LJ diameter) and left unshifted. The chemical potential $\mu$ is tuned to the conditions of bulk liquid-vapor coexistence: the (reduced) $\mu^*=-3.44610$\cite{Wilding1995} at a (reduced) temperature $T^*=1.0=0.842T_c$. The fluid interacts with both walls through a wall-fluid potential that is infinitely repulsive at $z=0$ and $z=L$ and has a long-range attraction of the modified $9$-$3$ LJ form previously studied by Evans {\em et al.} \cite{EvansStewartWilding2017}. By making the dimensionless wall-fluid attraction very weak (we chose $\epsilon_w=0.01$, in the notation of eq.~5 of ref.\cite{EvansStewartWilding2017}) we render the wall strongly solvophobic, giving rise to a state that is very close to critical drying \cite{EvansStewartWilding2017}. For such a state one expects substantial depletion of the density at each wall and greatly enhanced density fluctuations. Note that in our slit setup, the liquid is metastable w.r.t.~capillary evaporation, but this does not prevent us from studying the near-drying region at the walls (see the DFT study in Fig.~8 of ref.~\cite{EvansStewart2015} and Fig.~6 of ref.~\cite{EvansStewartWilding2017}).

\begin{figure}[t]
 \centering
 \includegraphics{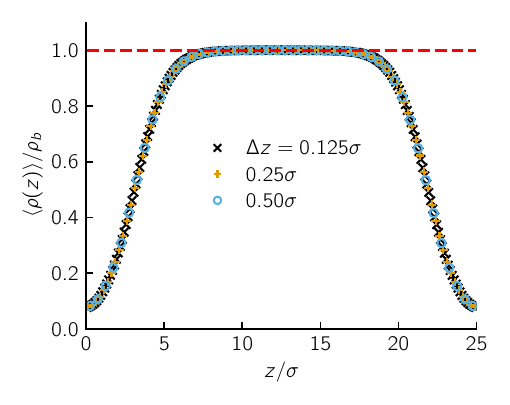}
 \caption{Monte Carlo results for the normalized density profile $\langle\rho(z)\rangle/\rho_b$ calculated for a truncated LJ liquid at bulk liquid-vapor coexistence and reduced temperature $T^*=1.0$ confined by a pair of solvophobic planar walls. Results are shown for three choices of subvolume size $V(z)=L^2\Delta z$ with $L=25\sigma$ and $\Delta z$ given in the legend. The overall system volume is $V=25\sigma^3$ and the dashed red line indicates $\langle \rho(z)\rangle =\rho_b$ where the bulk value is determined from an independent simulation of a large fully periodic system. Statistical errors are smaller than the symbol sizes.}
 \label{fig:ljslitden}
\end{figure}

We discretize the space between the walls into $L/\Delta z$ contiguous planar slab subvolumes of prescribed equal thickness $\Delta z$ and volume $V(z)=L^2\Delta z$. Histograms of the instantaneous density profile are formed as $\rho(z)=N(z)/V(z)$. Fig.~\ref{fig:ljslitden} shows our estimates of the ensemble average $\langle\rho(z)\rangle $ (normalized by the independently determined average bulk liquid density $\rho_b=0.654(1)$) for various choices of $\Delta z$. The principal features of these profiles are a considerable depletion in density near the solvophobic walls reflecting the incipient drying regions, and a relaxation to the bulk liquid density far from the walls. It should also be noted that within statistical uncertainties the profiles are independent of $\Delta z$.

\begin{figure}[t]
 \centering
 \includegraphics{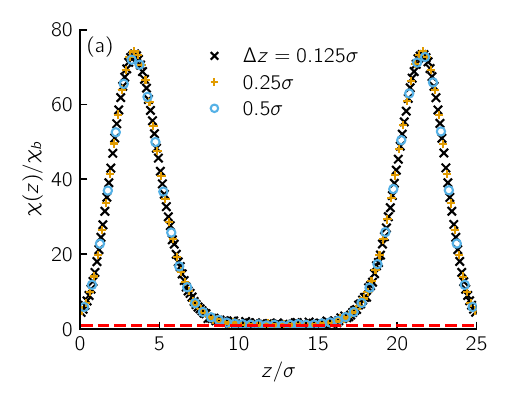}
 \includegraphics
 {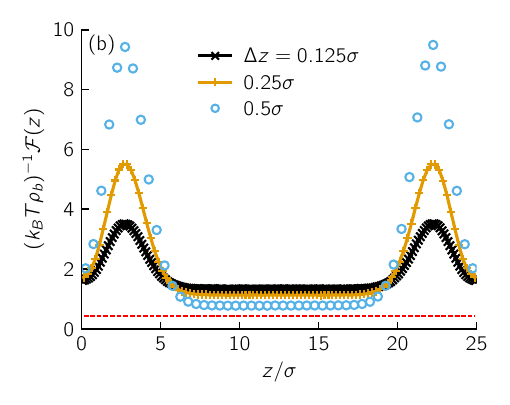}
 \caption{ ${\bf (a)}$ Monte Carlo results for the local compressibility profile $\chi(z)$ normalized by its bulk value $\chi_b$. These correspond to the density profile shown in Figure 1. Results are shown for three choices of subvolume size $V(z)=L^2\Delta z$ with $L=25\sigma$ and $\Delta z$ given in the legend. The overall system volume is $V=25\sigma^3$ and the dashed red horizontal line corresponds to $\chi(z)/\chi_b=1$. ${\bf (b)}$ The scaled fluctuation profile $(k_BT\rho_b)^{-1}{\cal F}(z)$ evaluated for the same subvolumes considered in {\bf (a)}.  Note the strong dependence on the subvolume thickness $\Delta z$ and that the profile does not decay in the slit middle to the correct value of the bulk compressibility $\kappa_T=0.432(3)$ in reduced LJ units,  as indicated by the dashed red horizontal line. In all cases, statistical errors are smaller than the symbol sizes.
 }
 \label{fgr:ljslitcompare}
\end{figure}

We now present our measurements of the local compressibility profile $\chi(z)$ and the spatial fluctuation profile ${\cal F}(z)$. The definitions and methodologies for calculating these quantities have been described in Sec.~\ref{sec:background}.  Results for $\chi(z)$ are shown in Fig.~\ref{fgr:ljslitcompare}(a) from which one sees (as already established in ref.~\cite{EvansStewartWilding2017}) that $\chi(z)$ exhibits strong enhancement --in this case by a factor of $~75$ compared to the bulk-- in the region of the incipient vapor-liquid interface that forms near the slit walls. Far from the walls, $\chi(z)$ decays precisely to its bulk value $\chi_b=\rho_b^2\kappa_T$ as denoted by the dashed red line ($\kappa_T=0.432(3)$ being determined by an independent simulation of the liquid in a large periodic system).  Similarly to $\langle \rho(z)\rangle$, we see that the profiles for $\chi(z)$ are essentially independent of the choice of $\Delta z$.

Matters are quite different for the spatial fluctuation profile ${\cal F}(z)$ shown in Fig.~\ref{fgr:ljslitcompare}(b) and scaled by a factor of $(k_BT\rho_b)^{-1}$ to allow comparison with the bulk compressibility $\kappa_T$ whose value is indicated by the red horizontal dashed line. ${\cal F}(z)$ was calculated using the same sequence of configurations and slab subvolume sizes as for $\chi(z)$. Nonetheless, the profile characteristics are quite distinct. Principally there is a strong dependence on subvolume thickness $\Delta z$. Although all profiles exhibit peaks near the wall, these are narrower than those of $\chi(z)$ and grow with increasing $\Delta z$.  The peak-to-trough enhancement factor for ${\cal F}(z)$ is considerably less than for $\chi(z)$, showing that the former is much less sensitive to variations in local density fluctuations than the latter. It is also less accurate: the profile for $(k_BT\rho_b)^{-1}{\cal F}(z)$ fails to decay to its bulk value $\kappa_T$ far from the walls. This latter observation accords with the bulk studies of Villamaina and Trizac~\cite{Villamaina_2014}. While further increasing the slab subvolume thickness $\Delta z$ would help to ameliorate these issues with ${\cal F}(z)$, this would come at the cost of a reduction in the resolution of the peak in this measure of density fluctuations. 

\subsection{Fluctuations at a spherical solvophobic solute investigated with spherical shell subvolumes}

\label{sec:sphere_comparison}

Next, we consider the case of a solvophobic spherical solute immersed in the same LJ solvent used in sec.~\ref{sec:slitcomparison}. The center of the solute particle is fixed at the origin and for the solute-solvent interaction we employ a potential of the form given in Eq.~(3) of Coe {\em et al} \cite{CoeEvansWildingJCP2023}, with the attractive well depth set to the very small (reduced) value $\epsilon_{sf}=0.01$ to render the solute strongly solvophobic. We partition the volume around the solute into concentric spherical shell subvolumes (centered on the solute) extending from radius $r$ to $r+\Delta r$. The instantaneous radial density profile is measured as $\rho(r)=N(r)/V(r)$ with $V(r)=4\pi r^2\Delta r$.  

\begin{figure}[t]
\centering
\includegraphics{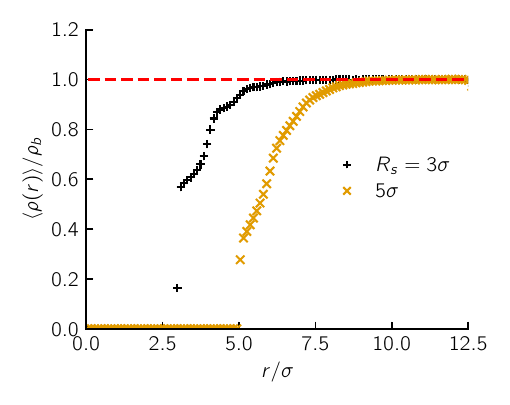}
\caption{Monte Carlo results for the radial density profiles for the truncated LJ fluid at liquid-vapor coexistence for reduced temperature $T^*=1.0$. The LJ solvent is in contact with a very weakly attractive spherical solute and results are shown for solute radii $R_s=3.0\sigma$ and $R_s=5.0\sigma$.  The radial profiles were accumulated using spherical shell subvolumes of thickness $\Delta r=0.125\sigma$. The overall system volume is $V=25\sigma^3$ and the dashed red horizontal line indicates $\rho_b$, the bulk value. Note the structure (`kinks') in the density profiles which stems from packing effects. Statistical errors are smaller than the symbol sizes.}
\label{fgr:ljsphdenscompare}
\end{figure}

In Fig.~(\ref{fgr:ljsphdenscompare}) we show $\langle \rho(r)\rangle$ normalized by its bulk value $\rho_b$ for two different solute particles of radii $R_s=3\sigma$ and $R_s=5\sigma$. The profiles were accumulated using spherical shells of equal thickness $\Delta r=0.0125\sigma$. One sees that in both cases close to the solute particle there is a pronounced depletion in solvent density reflecting the strong solvophobicity. The density profile in this region also exhibits `kinks', particularly for the smaller solute. These occur on length scales of about $\sigma$ and are remnants of the packing structure that occurs in bulk fluids. Away from the solute, the density relaxes to its bulk value $\rho_b$ as indicated by the red dashed horizontal line.

\begin{figure}[h]
\centering
\includegraphics{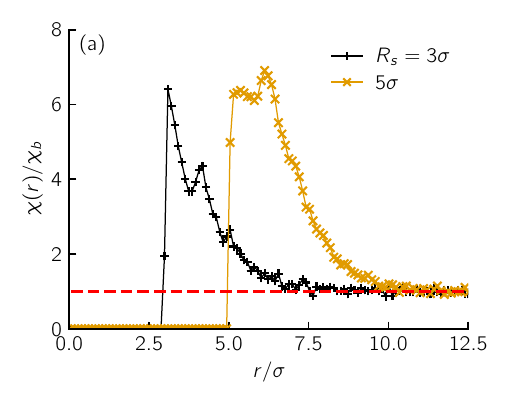}
\includegraphics{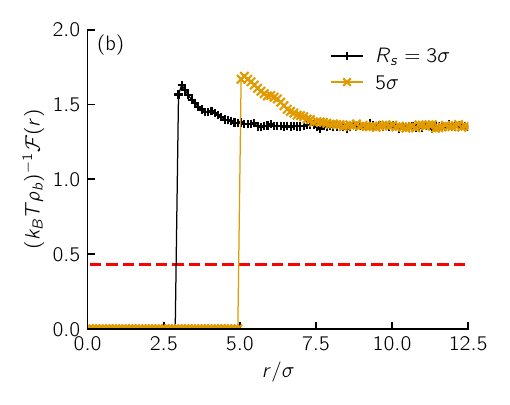}
\caption{{\bf (a)} Monte Carlo results for the radial local compressibility profile $\chi(r)$ for the truncated LJ fluid at liquid-vapor coexistence at $T^*=1.0$ and in contact with a very weakly attractive spherical solute of radius $R_s$. Results are shown for $R_s=3\sigma$ and $R_s=5\sigma$. The subvolume shell thickness is $\Delta r=0.125\sigma$ and the dashed red horizontal line corresponds to the bulk value. Note the structure in $\chi(r)$ which is related to the kinks in $\rho(r)$ seen in fig.~\ref{fgr:ljsphdenscompare}. {\bf (b)} Corresponding results for the scaled radial fluctuation profile ${\cal F}(r)$ for the same $\Delta r=0.125\sigma$. This profile resolves neither the extended range of enhanced density fluctuations seen in $\chi(r)$ nor the detailed structure. Moreover, it does not decay to the correct value of the bulk compressibility $\kappa_T=0.432(3)$ in reduced LJ units, as indicated by the dashed red horizontal line. In all cases, statistical errors are comparable with the symbol sizes.}
\label{fgr:ljsph}
\end{figure}

The corresponding local compressibility profiles $\chi(r)$ for the two solute radii are shown in Fig.~(\ref{fgr:ljsph})a. Close to the solute there is a great enhancement of the local compressibility compared to its bulk value. The peak in $\chi(r)$ for the larger solute particle extends over a greater range of radii than that for the smaller one, in accord with previous findings~\cite{CoeEvansWildingJCP2023}. Evident also is considerable structure in the profile. Whilst this is inherited from the density profile \cite{EvansStewartWilding2017} the structure is much richer and more pronounced, with subsidiary peaks in evidence. Far from the solute particle, the local compressibility decays smoothly and accurately to its bulk value $\chi_b$.

The corresponding scaled spatial fluctuation profiles are shown in Fig.~(\ref{fgr:ljsph})b, and were accumulated using the same subvolumes as for $\chi(r)$. Here one sees that ${\cal F}(r)$ exhibits a much weaker response to the enhanced fluctuations than does $\chi(r)$, and as in the slit case, the scaled profile fails to decay to the value of the bulk compressibility. Here the failure is dramatically clear.

\begin{figure}[h]
 \centering
 \includegraphics{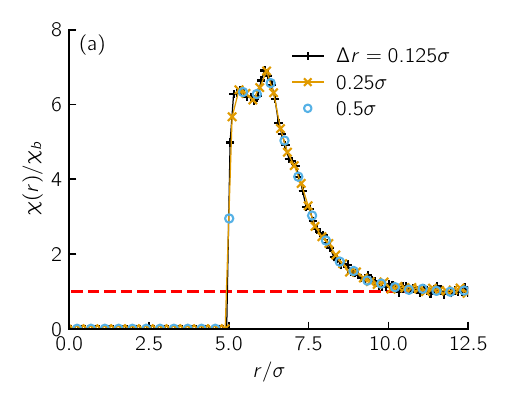}
 \includegraphics{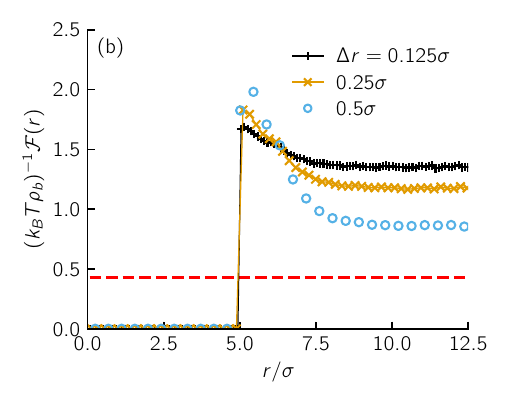}
 \caption{{\bf (a)} Monte Carlo results for the radial local compressibility profile $\chi(r)$ for the truncated LJ fluid at liquid-vapor coexistence and $T^*=1.0$ in contact with a very weakly attractive spherical solute of radius $R_s=5.0
 \sigma$. The dashed red horizontal line corresponds to the bulk value. Data are shown for three values of the spherical shell subvolume thickness as indicated in the legend. ${\bf (b)}$ The scaled radial fluctuation profile $(k_BT\rho_b)^{-1}{\cal F}(r)$ evaluated for the same subvolumes considered in (a).  Note the strong dependence on the subvolume shell thickness $\Delta r$ and that far from the solute the profile does not decay to the correct value of the bulk compressibility $\kappa_T=0.432(3)$ (in reduced LJ units), as indicated by the dashed red horizontal line. In all cases, statistical errors are comparable with the symbol sizes.}
 \label{fgr:ljsphcomparebins}
\end{figure}

To investigate the effects of varying the subvolume shell thickness on the results, we present in fig.~\ref{fgr:ljsphcomparebins} a comparison of profiles for $\chi(r)$ and ${\cal F}(r)$ for three different values of $\Delta r$. Fig.~\ref{fgr:ljsphcomparebins}(a) shows that $\chi(r)$ is insensitive to the subvolume size, and the detailed non-trivial structure of $\chi(r)$ is reproduced consistently for each $\Delta r$. By contrast the fluctuation profile ${\cal F}(r)$, Fig.~\ref{fgr:ljsphcomparebins}(b) proves to be very much less sensitive to the enhanced fluctuations near the solute surface than $\chi(r)$. Specifically, ${\cal F}(r)$  has a much smaller peak compared to its limiting value, fails to reflect accurately the range of the enhanced density fluctuations, and does not resolve the structural features picked up by $\chi(r)$. Whilst the sensitivity of ${\cal F}(r)$ increases with the subvolume size, this is at the cost of spatial resolution and does not approach the level of detail that is provided by $\chi(r)$.

\begin{figure}[h!]
 \centering
 \includegraphics{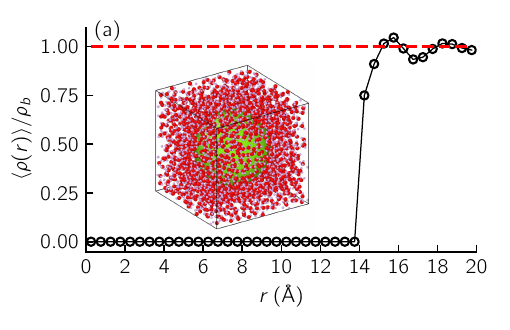}
 \includegraphics{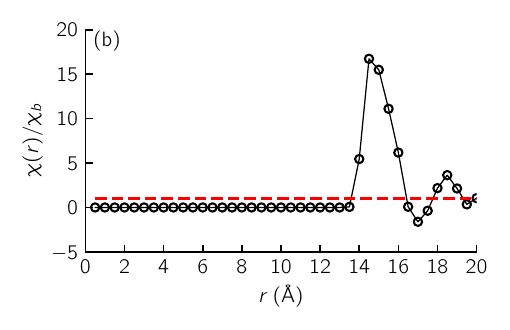}
 \includegraphics{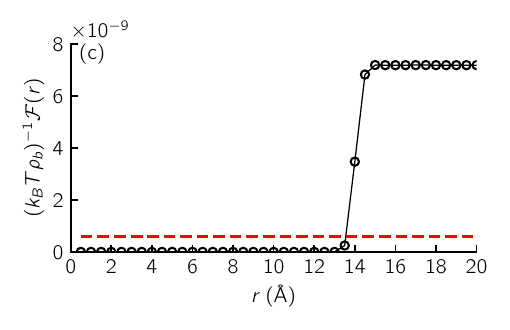}
 \caption{Monte Carlo results for SPC/E water in contact with a very weakly attractive spherical solute of radius $R_s=14$\AA~in a periodic box of volume $V=(40$\AA$)^3$. The temperature $T=300$K and $\mu$ is set to its bulk coexistence value \cite{NISTSPCE}. ({\bf a}) The radial density profile $\langle\rho(r)\rangle/\rho_b$; the inset shows a configurational snapshot. ({\bf b}) The local compressibility profile $\chi ( r)$. ({\bf c}) The scaled radial fluctuation profile ${\cal F}(r)$ in units of ${\rm m}^3{\rm J}^{-1}$.  Each of $\langle \rho(\bf r)\rangle$, $\chi ({\bf r})$ and ${\cal F}({\bf r})$ was calculated for a 3D lattice of cubic subvolumes of size $\Delta V=(0.5$\AA$)^3$, and then averaged spherically to produce the radial profiles shown. Statistical errors are comparable with the symbol sizes. The dashed red horizontal line in ({\bf a}) and ({\bf b}) corresponds to the bulk value; that in ({\bf c}) corresponds to the bulk compressibility $\kappa_T=5.92\times 10^{-10}$ m$^3$J$^{-1}$. Note the lack of response of ${\cal F}( r)$ to enhanced density fluctuations near the hydrophobic solute.}
 \label{fgr:SPCEsphcompare}
\end{figure}

\subsection{Fluctuations at a hydrophobic spherical solute investigated with cubic subvolumes for SPC/E water}

In this third example, we compare the local compressibility profile and the spatial fluctuation profile in cubic subvolumes. This is a scenario that one would likely adopt when seeking to map enhanced density fluctuations near an irregularly shaped entity that exhibits hydrophobic regions, as is the case for some large biomolecules.

The model that we have studied to illustrate this case is SPC/E water at a spherical solute, which we have simulated in the GCE using the open-source multipurpose Monte Carlo simulation engine DL\_MONTE\cite{dlmonte,dlmontesource}. The temperature was set to $T=300$K and the chemical potential to its corresponding coexistence value $\beta\mu=-15.24$ \cite{NISTSPCE}. A solute of radius $R_s=14$\AA~was fixed at the origin of a simulation box of size $V=(40$\AA$)^3$ and the solute-solvent interactions were assigned similarly to Sect.~\ref{sec:sphere_comparison} by employing a potential of the form given in eq.(3) of Coe {\em et al} \cite{CoeEvansWildingJCP2023} between the solute particle and the oxygen atom of the water molecules, with the attractive well depth fixed, as in the previous example, by the very small (reduced) value $\epsilon_{sf}=0.01$. The system was discretized into equal cubic subcells of size $V({\bf r})=(0.5$\AA$)^3$ and both the local compressibility profile $\chi({\bf r})$ and the spatial fluctuation profile ${\cal F}({\bf r})$ were accumulated w.r.t.~these. Although the subcells are the fundamental units on which measurements were performed, we can improve statistical sampling and generate a one-dimensional profile for comparison purposes by exploiting the symmetry of our system to average both $\chi({\bf r})$ and ${\cal F}({\bf r})$ spherically, yielding profiles for $\chi(r)$ and ${\cal F}(r)$. We stress, however, that the latter are not to be confused with the profiles that would have resulted from choosing spherical shell subvolumes in the manner of Sec.~\ref{sec:sphere_comparison}.

The density profile for this system is shown in Fig.~\ref{fgr:SPCEsphcompare}(a) and displays a $\sim 30\%$ depletion in the water density close to the solute and a weak oscillatory decay to the bulk. The comparison of the spherically averaged forms of $\chi({\bf r})$ and ${\cal F}({\bf r})$ is shown in Fig.~\ref{fgr:SPCEsphcompare}(b) and (c) respectively. In the former case, a strong enhancement of local compressibility is seen, the spatial range of which correlates with the depletion in the water density close to the solute, and which decays in a similar oscillatory fashion to the bulk value far from the surface. By contrast ${\cal F}({\bf r})$, while calculated in the same simulation and with the same set of subcells as used to calculate $\chi({\bf r})$, shows no signal of enhanced fluctuations at all --the spherically averaged profile $(k_BT\rho_b)^{-1}{\cal F}(r)$ rises from zero to a constant value which is approximately $12$ times that of the bulk compressibility $\kappa_T=5.92\times10^{-10}$m$^3$J$^{-1}$. This complete insensitivity of ${\cal F}({\bf r})$ to the density fluctuations is of course traceable to the small cell size $V({\bf r})=(0.5$\AA$)^3$ used in this case. While  Fig.~\ref{fgr:SPCEsphcompare}(b) shows that this choice of cell size is necessary and warranted to resolve the pertinent features of the local compressibility profile $\chi({\bf r})$, it results in a complete lack of signal in ${\cal F}({\bf r})$.

One can better understand this finding by considering the statistics of very small subvolumes\cite{Villamaina_2014}. When $V({\bf r})$ is smaller than the particle size, cell occupancy is limited to $N({\bf r})=0$ or $1$. Accordingly, the probability distribution function for occupation is binomial with some mean occupancy $\langle N({\bf r})\rangle$. Given this, one readily finds that the fluctuation profile eq.~\ref{eq:fluct_field} evaluates as ${\cal F}({\bf r})=1-\langle N({\bf r})\rangle$. When the subvolume cells are considerably smaller than the particle size or the solvent density is low, as can be the case in proximity to an extended solvo/hydrophobe, then $\langle N({\bf r})\rangle\ll 1$ and the statistics approach the Poissonian limit for which the variance and mean of the fluctuations in $N({\bf r})$ are equal. Accordingly, ${\cal F}({\bf r})$ approaches unity, which explains the absence of its response to density fluctuations as reflected in Fig.~\ref{fgr:SPCEsphcompare}(c). For position vectors ${\bf r}$ corresponding to bulk liquid SPC/E water, we find $\langle N({\bf r})\rangle=0.0041(1)$ which is consistent with the limiting (large \textit{r})  value of $(k_BT\rho_b)^{-1}{\cal F}({\bf r})=(k_BT\rho_b)^{-1}(1-\langle N({\bf r})\rangle)$ shown in Fig.~\ref{fgr:SPCEsphcompare}(c) given the measured value of the bulk molecular number density $\rho_b=3.348\times 10^{28}$m$^{-3}$.

\section{Summary and discussion}
\label{sec:discussion}

In this paper, we have compared the utility of two distinct approaches for measuring the strength of local density fluctuations in fluids near solvo/hydrophobes. The local compressibility $\chi({\bf r})$  introduced by Evans and Stewart~\cite{EvansStewart2015} in this context has been shown to provide high levels of sensitivity, resolution, and accuracy regardless of the geometry or length scales (subvolume size) for which it is applied. In particular as evidenced by Figs.~\ref{fgr:ljsph}(a) and \ref{fgr:SPCEsphcompare}(b),  $\chi({\bf r})$ can readily resolve the detailed features of the local compressibility even when they occur on sub-particle length scales.  As mentioned in the Introduction, this measure is well rooted in the statistical physics of interfacial phenomena, and its behavior in the vicinity of surface phase transitions is well established. 

By contrast, the spatial fluctuation profile ${\cal F}({\bf r})$, previously employed by several authors to study density fluctuations in simulations of water near hydrophobes, appears to be a considerably inferior measure in all respects. Specifically, it generally exhibits a much weaker response to the magnitude of local density fluctuations than $\chi({\bf r})$. This can only be mitigated by increasing the subvolume size, which comes at the cost of a loss of spatial resolution. The severity of this trade-off depends somewhat on the geometry. For a planar solvo/hydrophobic substrate the subvolumes that one can reasonably employ may not be so small as to render inadequate the sensitivity and resolution of the spatial fluctuation profile. However, for spherical solvo/hydrophobic solutes where the subvolumes in proximity to the surface are necessarily small, the sensitivity and resolution are considerably worse than those provided by the local compressibility. If one attempts to map the spatial fluctuations around an irregular hydrophobe such as a bio-molecule with acceptable resolution, then the spatial fluctuation profile will likely fail to provide an adequate signal of the local density fluctuations, while the local compressibility retains its full utility\footnote{Magnetic analogues of $\chi(z)$ and ${\cal F}(z)$ have been defined by Binder and coworkers to describe surface phase transitions in lattice-based magnetic models and termed $\chi_n$ and $\chi_{nn}$ respectively. In accordance with our study,  a much stronger magnetic fluctuation signal occurs for $\chi_n$ than for $\chi_{nn}$  \cite{Binder1995}. However,  in contrast to fluids, the choice of subvolume size is uniquely predefined in the lattice context.}

The differences between the two methods are traceable -in part at least- to the degree to which they satisfy the central limit theorem in the limiting case of a bulk fluid, as discussed in Sections~\ref{sec:background} and \ref{sec:MCresults}. On the smallest length scales the occupation statistics of subvolumes are binomial, tending to the Poissonian limit. Gaussian density fluctuations emerge in the bulk only for sufficiently large subvolumes whose linear extent greatly exceeds the local correlation length, and which contain correspondingly large numbers of particles.  Comparing the correlators that define the two approaches, eqs.~(\ref{eq:lc_corr2}) and (\ref{eq:fluct_field}), one sees that the typical magnitude of the product $\langle N({\bf r})N\rangle$ in the correlator for $\chi({\bf r})$ far exceeds that of $\langle N^2({\bf r})\rangle$ appearing in ${\cal F}({\bf r})$.  It follows that the linear scaling, with subvolume size, of the numerator in the correlator that is required to yield an intensive measure of fluctuation strength is readily achieved for $\chi({\bf r})$ but not  for $ {\cal F}({\bf r})$. This is why $\chi({\bf r})$ yields accurate estimates of the bulk compressibility irrespective of the choice of subvolume size. And these benefits of accuracy and high resolution extend beyond the case of purely bulk fluctuations to yield greater sensitivity to the non-Gaussian near-critical fluctuations that are the root cause of enhanced density fluctuations near extended solvo/hydrophobes \cite{CoeEvansWildingJCP2023}.

While the spatial fluctuation profile ${\cal F}({\bf r})$ has been employed in several papers to study enhanced density fluctuation near extended hydrophobes \cite{WillardChandler2014,AcharyaGarde2010,SarupriaGarde2009}, we note that others \cite{PatelVarillyChandler2010,Jamadagni2011,rego2022} refer instead to the form of the pdf of the subvolume particle number $P(N_v)$. The appearance of enhanced fluctuations in the form of a non-Gaussian tail at small $N_v$ has been reported and attempts made to relate such behavior to the degree of hydrophobicity of the substrate/ solute~\cite{PatelVarillyChandler2010}. However, this tail is apparently visible only for quite large subvolumes \cite{rego2022}, which seems to rule out the use of $P(N_v)$ to create a high-resolution fluctuation profile.

Finally, we note that a key feature of the local compressibility $\chi({\bf r})$ is that it is defined within the GCE. This open ensemble is doubtlessly optimal for studying density fluctuations in inhomogeneous systems because fluctuations can occur on all length scales up to and including the system size. Furthermore, the GCE lends itself to accurate positioning of the thermodynamic state of interest relative to (bulk) phase coexistence. Recall that the deviation from bulk liquid-vapor coexistence, measured by the chemical potential deviation, is a crucial ingredient in ascertaining the origin, range and strength of solvo/hydrophobicity-induced density fluctuations \cite{EvansStewartWilding2017}. Additionally, as we have seen, the GCE offers practical efficiencies such as the ability to calculate  $\chi({\bf r})$ via histogram reweighting (see Sec.~\ref{sec:LCPdef}). Grand Canonical Monte Carlo is implemented in several general-purpose molecular simulation engines such as DL\_MONTE\cite{dlmontesource,dlmonte} and LAMMPS~\cite{LAMMPS} and is thus readily accessible for a wide range of applications including complex molecules. Nevertheless, it must be acknowledged that many simulation studies of hydrophobic phenomena are not performed grand canonically but rather using molecular dynamics in a closed (constant-$N$) ensemble. While we expect the deficiencies of the spatial fluctuation profile that we have identified in the GCE to be at least as pronounced in closed ensembles as they are in the GCE, future work could usefully consider the benefits offered by analogs of the local compressibility to closed ensembles. Significant steps in this direction have recently been reported by Eckert et al~\cite{Eckert_2023} who consider their local thermal susceptibility in both the canonical and grand ensembles.

\vspace*{2mm}

%Possible additions:

%\begin{itemize}

%\item Should we say more about the criticality eg. point out existence of the large parallel correlation length in the planar case and that even $\chi(z)$ won't converge until the slab length $L\gg\xi_\parallel$? Maybe in the introduction?

%\item Hummer et al Proc. Natl. Acad. Sci. USA Vol. 93, pp. 8951-8955, August 1996 claims Gaussian fluctuations for spherical subvolumes with a mean of 10 water molecules. Pedersen \cite{Pedersen2019} says something similar for 20 LJ molecules. Note Fig 7 of Pedersen \cite{Pedersen2019} for a LJ liquid in the absence of a hydrophobe also shows some signs of a tail to low density even though there is no hydrophobe, so need to be careful interpreting eg fig 1d of Rego et al\cite{rego2022}. Could point out that a more accurate criterion for whether the fluctuations are truly Gaussian may not be the quality of a fit, but rather whether the variance scales correctly with subvolume size? This is probably  out of scope. 

%\end{itemize}
\vspace*{2mm}
\acknowledgements
NBW is grateful to Tom Underwood for guidance in the use of DL\_MONTE. The computer simulations were carried out using the computational facilities of the Advanced Computing Research Centre, University of Bristol, as well as the Isambard 2 UK National Tier-2 HPC Service (\href{http://gw4.ac.uk/isambard/}{http://gw4.ac.uk/isambard/}) operated by GW4 and the UK Met Office, and funded by EPSRC (EP/T022078/1). RE acknowledges support of the Leverhulme Trust, grant no. EM-2020-029$\backslash$4

\appendix
\section{Correlator for the local compressibility}
\label{sec:appendcorr}

For completeness, we include a derivation of Eq.~\ref{eq:lc_corr}
In the grand canonical (constant-$\mu,V,T$) ensemble, the particle number $N$ fluctuates under the control of the chemical potential $\mu$. Denote a configuration of $N$ particles as $\{r_i\}^N$ where $i=0\ldots N$. For an inhomogeneous fluid under the influence of an external potential $V_{ext}({\bf r})$ that is a function of the position vector ${\bf r}$, the Hamiltonian can be written $H(\{r_i\}^N|\mu,V,T)=E(\{r_i\}^N)+\int d{\bf r}\rho({\bf r})V_{ext}({\bf r})-\mu N$, where $E(\{r_i\}^N)$ is the configurational energy. The grand partition function follows as $\Xi(\mu,V,T)=\sum_{N=0}^\infty\int (\prod_{i=0}^{N} dr_i)e^{-\beta H}$, with $\beta=(k_BT)^{-1}$where we have suppressed phase space and combinatorical factors.

The average density profile is obtained by a functional derivative of the grand potential $\Omega=-\beta^{-1}\ln\Xi$ w.r.t.~the external potential  

\begin{eqnarray}
\frac{\delta \Omega}{\delta V_{ext}({\bf r})}&=& -\frac{1}{\beta\Xi}\frac{\delta \Xi}{\delta V_{ext}({\bf r})}\\
&=&\frac{1}{\beta\Xi}\sum_{N=0}^\infty\int (\prod_{i=0}^{N} dr_i)\beta\rho({\bf r})e^{-\beta H} \\
&=&\langle\rho({\bf r})\rangle\:,
\end{eqnarray}
which is a standard result. 

Now define the local compressibility profile
\begin{widetext}
\begin{eqnarray}
    \chi({\bf r}) &\equiv& \left.\frac{\partial \langle \rho({\bf r})\rangle}{\partial \mu}\right|_T\\
     &=& \sum_{N=0}^\infty\int (\prod_{i=0}^{N} dr_i)\rho({\bf r})\left[\frac{1}{\Xi}\beta Ne^{-\beta H}+e^{-\beta H}\frac{1}{\Xi^2}\frac{\partial \Xi}{\partial\mu}\right].\\
     \end{eqnarray}
     Thus
     \begin{eqnarray}
   k_BT\chi({\bf r})&=&\sum_{N=0}^\infty\int (\prod_{i=0}^{N} dr_i)\left[N\rho({\bf r})\frac{e^{-\beta H}}{\Xi}-\rho({\bf r})\frac{e^{-\beta H}}{\Xi} \langle N\rangle\right]\\[2mm]
        &=&\langle\rho({\bf r}) N\rangle-\langle\rho({\bf r})\rangle \langle N\rangle\:.
        \label{eq:muVTcorrelator}
\end{eqnarray}
\end{widetext}

This result simplifies in the case of a bulk (uniform) system having constant number density $\rho_b$, for which $\rho({\bf r})=\rho_b=\langle N\rangle/V$ and $\chi({\bf r})=\chi_b$.  Integrating eq.~(\ref{eq:muVTcorrelator}) w.r.t.~the differential volume $dV$, over the fixed system volume and noting that $\int_0^V dV=V$ and $\int_0^V \rho({\bf r})dV=N$, one obtains

\begin{eqnarray}
Vk_BT \chi_b&=&\langle N^2\rangle-\langle N\rangle^2\\[2mm]
\frac{Vk_BT \chi_b}{\langle N\rangle}&=&\frac{\langle N^2\rangle-\langle N\rangle^2}{\langle N\rangle}= \rho_b k_BT\kappa_T\:,
\end{eqnarray}
where we have used Eq.~(\ref{eq:kappaT}).  It follows that in the bulk 

\begin{equation}
    \chi_b=\rho_b^2\kappa_T.
\end{equation}

\bibliography{references}% Produces the bibliography via BibTeX.
\end{document}